\documentclass[12pt]{article}

\usepackage[T1]{fontenc}
\usepackage[latin9]{inputenc}
\usepackage{color}
\usepackage{amsmath}
\usepackage{amssymb}
\usepackage{graphicx}
\usepackage[authoryear]{natbib}

\usepackage{amsmath}
\usepackage{graphicx,psfrag,epsf}
\usepackage{enumerate}
\usepackage{multirow}
\usepackage{url} 
\usepackage{graphicx}
\usepackage{color}
\usepackage{float}
\usepackage{mathrsfs}
\usepackage{amsmath}
\usepackage{amssymb}

\usepackage{graphicx}
\usepackage{esint}
\usepackage{latexsym}
\usepackage[authoryear]{natbib}
\usepackage{times}
\usepackage{bm}
\usepackage{amsfonts}
\usepackage{multirow}
\usepackage{graphicx}
\usepackage{caption}

\newcommand{\biblist}{\begin{list}{}
{\listparindent 0.0cm \leftmargin 0.50cm \itemindent -0.50 cm
\labelwidth 0 cm \labelsep 0.50 cm
\usecounter{list}}\clubpenalty4000\widowpenalty4000}
\newcommand{\ebiblist}{\end{list}}

\usepackage{latexsym}
\usepackage{multirow}

\newtheorem{theorem}{Theorem}
\newtheorem{assumption}{Assumption}
\newtheorem{remark}{Remark}
\newtheorem{proof}{Proof}

\newtheorem{definition}{Definition}

\newcommand{\be}{\begin{equation}}
\newcommand{\en}{\end{equation}}
\newcommand{\bea}{\begin{eqnarray}}
\newcommand{\ena}{\end{eqnarray}}
\newcommand{\ba}{\begin{array}}

\newcommand{\ea}{\end{array}}

\newcommand{\blind}{0}

\begin{document}

\def\spacingset#1{\renewcommand{\baselinestretch}%
{#1}\small\normalsize} \spacingset{1}


\if0\blind
{
  \title{\bf Propensity score weighting for causal inference with multi-stage
  	clustered data }
  \author{Shu Yang\\
		Department of Statistics, North Carolina State University}
	\date{}	
  \maketitle
} \fi

\if1\blind
{
  \bigskip
  \bigskip
  \bigskip
  \begin{center}
    {\LARGE\bf Propensity score weighting for causal inference with multi-stage
    	clustered data}
\end{center}
  \medskip
} \fi

\bigskip
\begin{abstract}
Propensity score weighting is a tool for causal inference to adjust
for measured confounders. Survey data are often collected under complex
sampling designs such as multi-stage cluster sampling, which presents
challenges for propensity score modeling and estimation. In addition,
for clustered data, there may also be unobserved cluster effects related
to both the treatment and the outcome. When such unmeasured confounders
exist and are omitted in the propensity score model, the subsequent
propensity score adjustment will be biased. We propose a calibrated
propensity score weighting adjustment for multi-stage clustered data
in the presence of unmeasured cluster-level confounders. The propensity
score is calibrated to balance design-weighted covariate distributions
and cluster effects between treatment groups. In particular, we consider
a growing number of calibration constraints increasing with the number
of clusters, which is necessary for removing asymptotic bias that
is associated with the unobserved cluster-level confounders. We show
that our estimator is robust in the sense that the estimator is consistent
without correct specification of the propensity score model. We extend
the results to the multiple treatments case. In simulation studies
we show that the proposed estimator is superior to other competitors.
We estimate the effect of School Body Mass Index Screening on prevalence
of overweight and obesity for elementary schools in Pennsylvania. 
\end{abstract}

\noindent%
{\it Keywords:}  Causality; Covariate Balance; Double Robustness; Unmeasured
Confounder.
\vfill

\newpage
\spacingset{1.45} 
\section{Introduction}
\label{sec:intro}

The gold standard for evaluating effects of treatments is using randomized
controlled trials. However, this approach may not be applicable due
to practical constraints or ethical issues. Observational studies
become useful in these settings. In observational studies, there often
is confounding by indication: some covariates are predictors of both
the treatment and the outcome. One implication is that the covariate
distributions differ between treatment groups. Under the assumption
of unconfoundedness or ignorable treatment assignment, causal effect
of treatments can be obtained by comparing the outcomes for units
from different treatment groups, adjusting for the observed confounders.
\citet{rosenbaum1983central} further introduced the central role
of the propensity score, and showed that adjusting for the propensity
score is sufficient to remove bias due to all observed confounders.
An extensive literature thereafter proposed a number of estimators
based on the propensity score. In particular, propensity score weighting
can be used to create a weighted population where the covariate distributions
are balanced between treatment groups, and the comparison between
the weighted outcomes has a causal interpretation \citep{hirano2001estimation,hirano2003efficient,imbens2015causal}. 

Survey data are observational in nature, and are often collected under
complex sampling designs. In complex surveys, each unit is associated
with a design weight, which approximates the number of units that
this unit represents in the finite population. Also, the data often
undergo other weighting adjustments such as calibration, nonresponse
adjustment, poststratification, raking, weight trimming, and etc.
Propensity score methods for causal inference are well-developed for
non-survey data; there is however much less literature that focuses
on how to adopt these methods for complex survey data, with exceptions
including \citet{zanutto2005using,zanutto2006comparison,li2013propensity},
and \citet{dugoff2014generalizing}. These researchers suggested that
design weights should be incorporated in propensity score modeling
or the weighting estimators. For clustered data, \citet{li2013propensity}
investigated the performance of the propensity score weighting estimator
and the doubly robust estimator under generalized linear mixed effect
models for the propensity score and the outcome, and showed that at
least either the propensity score model or the weighting estimator
should take the sampling design into account in order to avoid bias. 

Another challenge with complex survey data is that the sampled data
are often collected at multi-stages. For example, for two-stage cluster
sampling, clusters are selected at the first stage, and then for the
sampled clusters, units are selected at the second stage. The multi-stage
sampling design makes the propensity score modeling difficult. Moreover,
even we collect a rich set of unit-level covariates, there may be
unobserved cluster effects related to both the treatment and the outcome.
When such unmeasured confounders exist and are omitted in the propensity
score model, the subsequent analysis will be biased. 

The goal of this article is to develop propensity score weighting
for complex survey data with multi-stage clustered data structure
in the presence of unmeasured cluster-level confounders. We focus
on two-stage cluster sampling. The key insight is based on the central
role of the propensity score in balancing the covariate distributions
between treatment groups in the finite population. In survey sampling,
calibration is widely used to integrate auxiliary data, see for example,
\citet{chen1999pseudo}; \citet{wu2001model}; \citet{chen2002using};
and \citet{kim2009calibration}, or to handle nonresponse in survey
sampling, see for example, \citet{kott2006using}; \citet{chang2008using};
and \citet{kim2016calibrated}. In causal inference, calibration has
been used such as Constrained Empirical Likelihood \citep{qin2007empirical},
Entropy Balancing \citep{hainmueller2012entropy}, Inverse Probability
Tilting \citep{graham2012inverse}, and Covariate Balance Propensity
Score of \citep{imai2014covariate}. \citet{chan2015globally} showed
that estimation of average treatment effects by empirical balancing
calibration weighting can achieve global efficiency. However, all
these works are developed for simple settings with independent and
identically distributed (iid) random variables and they assume that
there are no unmeasured confounders. We adopt calibration for causal
inference with clustered data, to handle unmeasured cluster effects
that may confound the causal relationship between the treatment and
the outcome. Based on the sample, we impose the design-weighted covariate
balancing constraints, and also certain design-weighted balancing
constraints for each cluster. In particular, we consider a growing
number of calibration constraints increasing with the number of clusters,
which is necessary for removing asymptotic bias that is associated
with the unobserved cluster-level confounders. 

\textcolor{black}{The organization of this paper is as follows. Section
	2 provides the basic setup. Section 3 introduces the proposed calibration
	propensity score weighting estimator and the computational aspect
	in light of exponential titling. In Section 4, main results are presented.
	U}nder certain conditions, we show that the proposed estimator is
consistent for the average treatment effect in the presence of unmeasured
cluster-level confounders, without requiring correctly specification
of the propensity score model and the outcome model, and therefore
is robust. Imposing calibration conditions also improves the efficiency
of the estimator. In Section 5, we extend the results to the multiple
treatments case. In Section 6, we examine the consistency and robustness
of the proposed estimator in finite samples by simulation. In Section
7, we estimate the effect of School Body Mass Index Screening on prevalence
of overweight and obesity for elementary schools in Pennsylvania,\textcolor{black}{{}
	and discussions are made in Section 8.}

\spacingset{1.45} 
\section{Basic Setup}

We use the potential outcome framework \citep{rubin1974estimating},
which has been commonly adopted in the causal inference literature.
Consider a finite population with $M$ clusters and $N_{i}$ units
in the $i$th cluster. Therefore, the population size is $N=\sum_{i=1}^{M}N_{i}$.
For unit $j$ in cluster $i$, we observe a vector of pre-treatment
variables $X_{ij}$, a binary treatment $A_{ij}$ with 0 indicating
the control treatment and $1$ indicating the active treatment, and
lastly an outcome variable $Y_{ij}$. We assume that there is no interference
between units and no versions of each treatment level (the Stable
Unit Treatment Value assumption, \citealp{rubin1978bayesian}). Under
this assumption, each unit has two potential outcomes: $Y_{ij}(0)$,
the outcome that would be realized if the unit received the control
treatment, and $Y_{ij}(1)$, the outcome that would be realized if
the  unit received the active treatment. We assume that the observed
outcome is the potential outcome corresponding to the treatment received,
i.e., $Y_{ij}=Y_{ij}(A_{ij})$ (the Consistency assumption, \citealp{rubin1974estimating}). 

The sample is selected according to a two-stage cluster sampling design.
Specifically, at the first stage, cluster $i$ is sampled with the
first inclusion probability $\pi_{i}$, $i\in S_{I}$, where $S_{I}$
is the index set for the sampled clusters and we assume that $S_{I}=\{1,\ldots m\}$
for simplicity. Let $\pi_{ij}=\mathrm{pr}(i,j\in S_{I})$ be the second
inclusion probability for clusters $i$ and $j$ being sampled. At
the second stage, given that cluster $i$ was selected at the first
stage, unit $j$ is sampled with conditional probability $\pi_{j|i}$,
$j=1,\ldots,n_{i}$. Let $\pi_{kl|i}$ be the second inclusion probability
for units $k$ and $l$ being sampled given that cluster $i$ was
selected. The final sample size is $n=\sum_{i\in S_{I}}n_{i}$. Let
the design weight for unit $j$ in cluster $i$ be $\omega_{ij}=(\pi_{i}\pi_{j|i})^{-1}$,
which reflects the number of units for cluster $i$ in the finite
population this unit $j$ represents. Our goal is to estimate the
average treatment effect $\tau=E\{Y(1)-Y(0)\}$ based on the sample. 

\citet{rubin1974estimating} described the condition for estimating
average treatment effect in the setting with iid samples, the so-called
unconfoundedness or ignorable treatment assignment assumption, 
\begin{equation}
	Y(a)\bot A\mid X,\label{eq:NUC}
\end{equation}
for $a=0,1$. This assumption indicates that there are no unmeasured
confounders, which can be achieved by collected a sufficiently rich
set of pre-treatment variables that affect both the treatment and
the outcome. For clustered data, even we collect all the unit-level
confounders, there may be unmeasured cluster effects $U_{i}$ that
are related to both the treatment and the outcome. In this case, we
make the following assumption instead of (\ref{eq:NUC}).

\begin{assumption}[Ignorability]\label{assump1}For $a=0,1$, $Y_{ij}(a)\bot A_{ij}\mid X_{ij},U_{i}$.
	
\end{assumption} 

Under Assumption \ref{assump1}, 
\begin{equation}
	E\{Y_{ij}(a)\mid X_{ij},U_{i}\}=E(Y_{ij}\mid A_{ij}=a,X_{ij},U_{i}),\label{eq:outcome reg}
\end{equation}
so the average of the potential outcomes can be identified if the
cluster effects $U_{i}$ are observed. 

Following \citet{rosenbaum1983central}, we define the propensity
score for our setting. 

\begin{definition}[Propensity score] The propensity score is the
	conditional probability of receiving the active treatment given the
	confounders, 
	
	\begin{equation}
		e(X_{ij},U_{i})=\mathrm{pr}(A_{ij}=1\mid X_{ij},U_{i}).\label{eq:model2}
	\end{equation}
	
\end{definition} 

To estimate the average treatment effect, we make the following identifiable
assumptions. First, let us assume that there is sufficient overlap
between treatment groups. 

\begin{assumption}[Overlap]\label{assump2} For all $X_{ij}$ and
	$U_{i}$, there exist $\underline{e}$ and $\bar{e}$ such that $0<\underline{e}<e(X_{ij},U_{i})<\bar{e}<1$. 
	
\end{assumption} 

The above overlap assumption is required; otherwise there exist some
units for which we can not estimate the treatment effect without extrapolation
assumptions. Secondly, since the cluster effects are never observed,
we make the following independence assumption in order to identity
the causal parameter. 

\begin{assumption}\label{assump3-indepence}The unit-level confounder
	$X_{ij}$ and the unobserved cluster-level confounder $U_{i}$ are
	independent. 
	
\end{assumption} 

\begin{figure}
	\begin{centering}
		\includegraphics[width=0.8\columnwidth]{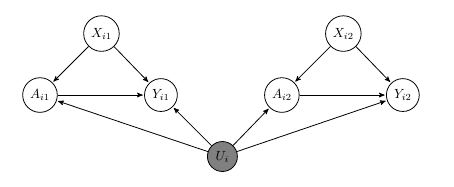}
		\par\end{centering}
	\caption{\label{fig:DAG}A Directed Acyclic Graph illustration for cluster
		$i$ with two units.}
\end{figure}

The unobserved cluster-level confounder $U_{i}$ can be viewed as
a modeling quantity, which is similar to the role of the random effect
in mixed effect models. In practice, the unobserved cluster-level
confounder is likely to be associated with the observed confounders.
In such cases, we model $U_{i}$ to be part of the unobserved confounder
that is independent of $X_{ij}$. Implicitly, we assume that the other
part of the unobserved confounder is fully controlled after adjusting
for $X_{ij}$. Figure \ref{fig:DAG} presents a causal diagram for
which Assumptions \ref{assump1} and \ref{assump3-indepence} are
satisfied. In Theorem \ref{Thm:unbiasedness}, we show that under
Assumptions \ref{assump1}\textendash \ref{assump3-indepence}, $\tau$
is nonparametrically identifiable. 

Notice that the outcome model for (\ref{eq:outcome reg}) and the
propensity score model for (\ref{eq:model2}) share the same random
effect $U_{i}$. Such models, the so-called shared parameter or shared
random effects models, have been used in the missing data literature
for modeling one particular type of nonignorable missingness. Namely,
researchers use separate models for the primary response and missingness
and link them by a common random parameter. See for example, \citet{follmann1995approximate,gao2004shared,Yang2013parametric,kim2016calibrated}. 

Finally, for the asymptotic result, we assume the following moment
condition. 

\begin{assumption}\label{assump4-moment}For $a=0,1$, $E\{Y(a)^{4}\}<\infty$.
	
\end{assumption} 

\spacingset{1.45} 
\section{Methodology }

For analyzing survey data, design-based approaches are favored in
government agencies since they are model-free and thereby avoiding
model misspecification \citep{kish1965survey,cochran2007sampling}.
The widely used design-based estimator is the Horvitz-Thompson estimator
\citep{horvitz1952generalization}. For example, let $T$ is the population
total of $Y_{ij}$, and the Horvitz-Thompson estimator is $\hat{T}_{n}=\sum_{i\in S_{I}}\sum_{j=1}^{n_{i}}\omega_{ij}Y_{ij}$,
which is design-unbiased for $T$. 

If the propensity score $e(X_{ij},U_{i})$ is known, the design-based
inverse probability of treatment weighting (IPTW) estimator for $\tau$
is 
\begin{equation}
	\hat{\tau}_{\mathrm{IPTW}}=\frac{1}{\hat{N}}\sum_{i\in S_{I}}\sum_{j=1}^{n_{i}}\omega_{ij}\left\{ \frac{A_{ij}Y_{ij}}{e(X_{ij},U_{i})}-\frac{(1-A_{ij})Y_{ij}}{1-e(X_{ij},U_{i})}\right\} ,\label{eq:ps estimator}
\end{equation}
where $\hat{N}=\sum_{i\in S_{I}}\sum_{j=1}^{n_{i}}\omega_{ij}$. Now
the issue is that in observational studies, the true propensity score
is usually unknown. For propensity score modeling and estimation,
the majority of the literature relies on parametric logistic regression
to estimate propensity score. For clustered data, \citet{li2013propensity}
considered different models including fixed effect logistic regression
models and random effect logistic regression models. However, it requires
assumptions regarding variable selection, the functional form of variables,
and specification of interactions. If any of these assumption fail,
it may results in bias in effect estimation. 

We consider a parametric working model for the propensity score, and
calibrate the propensity score to satisfy certain constraints. To
motivate these constraints, notice that the central role of the propensity
score is to balance the covariate distributions between treatment
groups in the population. Specifically, we have 
\begin{equation}
	E\left\{ \frac{A}{e(X,U)}X\right\} =E\left\{ \frac{1-A}{1-e(X,U)}X\right\} =E(X),\label{eq:C1-2}
\end{equation}
and 
\begin{equation}
	E\left\{ \frac{A}{e(X,U)}U\right\} =E\left\{ \frac{1-A}{1-e(X,U)}U\right\} =E(U).\label{eq:C3-4}
\end{equation}

Based on the sample, for the estimated propensity score $\hat{e}(X_{ij},U_{i})$,
we would impose the following design-weighted moment constraints,
\begin{eqnarray}
	\sum_{i\in S_{I}}\sum_{j=1}^{n_{i}}\omega_{ij}\frac{A_{ij}}{\hat{e}(X_{ij},U_{i})}X_{ij} & = & \sum_{i\in S_{I}}\sum_{j=1}^{n_{i}}\omega_{ij}X_{ij},\label{eq:c1}\\
	\sum_{i\in S_{I}}\sum_{j=1}^{n_{i}}\omega_{ij}\frac{1-A_{ij}}{1-\hat{e}(X_{ij},U_{i})}X_{ij} & = & \sum_{i\in S_{I}}\sum_{j=1}^{n_{i}}\omega_{ij}X_{ij},\label{eq:c2}\\
	\sum_{i\in S_{I}}\sum_{j=1}^{n_{i}}\omega_{ij}\frac{A_{ij}}{\hat{e}(X_{ij},u_{i})}U_{i} & = & \sum_{i\in S_{I}}\sum_{j=1}^{n_{i}}\omega_{ij}U_{i},\label{eq:c3*}\\
	\sum_{i\in S{}_{I}}\sum_{j=1}^{n_{i}}\omega_{ij}\frac{1-A_{ij}}{1-\hat{e}(X_{ij},u_{i})}U_{i} & = & \sum_{i\in S_{I}}\sum_{j=1}^{n_{i}}\omega_{ij}U_{i},\label{eq:c4*}
\end{eqnarray}
which approximate equations in (\ref{eq:C1-2}) and (\ref{eq:C3-4}).
However, since the cluster effects $U_{i}$ are unobserved, the constraints
(\ref{eq:c3*}) and (\ref{eq:c4*}) are infeasible. We notice that
if 
\begin{eqnarray}
	\sum_{j=1}^{n_{i}}\omega_{ij}\frac{A_{ij}}{\hat{e}(X_{ij},u_{i})} & = & \sum_{j=1}^{n_{i}}\omega_{ij},\ (i=1,\ldots,K),\label{eq:c3}\\
	\sum_{j=1}^{n_{i}}\omega_{ij}\frac{1-A_{ij}}{1-\hat{e}(X_{ij},u_{i})} & = & \sum_{j=1}^{n_{i}}\omega_{ij},\ (i=1,\ldots,K),\label{eq:c4}
\end{eqnarray}
the constraints (\ref{eq:c3*}) and (\ref{eq:c4*}) are satisfied
automatically. Here, we consider a growing number of calibration equations
increasing with the number of clusters, as opposed to a fixed number
of calibration equations (\ref{eq:c3*}) and (\ref{eq:c4*}). The
growing number of calibration equations is necessary for removing
asymptotic bias that is associated with the unobserved cluster-level
confounders. 

\subsection{Computation}

We now discuss the specific steps for computation. 
\begin{description}
	\item [{Step0.}] Use a logistic linear fixed effect model with a cluster-level
	main effect, fitted to $(A_{ij},X_{ij},\delta_{i})$ where $\delta_{i}$
	is the cluster indicator. This provides an initial set of inverse
	propensity score weights $\mathbb{W}^{0}=\{d_{ij};i\in S_{I},j=1,\ldots,n_{i}\}$,
	with $d_{ij}=1/e_{ij}^{0}$ if $A_{ij}=1$ and $d_{ij}=1/(1-e_{ij}^{0})$
	if $A_{ij}=0$.
	\item [{Step1.}] We follow the procedure discussed by\textcolor{black}{{}
		\citet{deville1992calibration}} to minimize a function of the distance
	between the initial weights $\mathbb{W}^{0}$ and the final weights
	$\mathbb{W}=\{\alpha_{ij};i\in S_{I},j=1,\ldots,n_{i}\}$, subject
	to the calibration constraints. Consider a general distance function
	\[
	\min\sum_{i\in S_{I}}\sum_{j=1}^{n_{i}}G(\alpha_{ij},\omega_{ij})
	\]
	subject to the calibration constraints. If $G(\alpha_{ij},\omega_{ij})=\omega_{ij}(\alpha_{ij}/\omega_{ij}-1)^{2}$,
	the minimum distance estimation leads to the generalized regression
	estimator \citep{park2012generalized}. If $G(\alpha_{ij},\omega_{ij})=-\omega_{ij}\log(\alpha_{ij}/\omega_{ij})$,
	this approach leads to empirical likelihood estimation \citep{newey2004higher}.
	Calibration using empirical likelihood has been discussed in \citet{hellerstein1999imposing};
	\citet{tan2006distributional}; \citet{qin2007empirical}; \citet{chan2012uniform};
	\citet{graham2012inverse}; \citet{han2013estimation}. We modify
	the initial set of weights $\mathbb{W}^{0}$ to a new set of weights
	$\mathbb{W}=\{\alpha_{ij};i\in S_{I},j=1,\ldots,n_{i}\}$ by minimizing
	the Kullback-Leibler distance \citep{kullback1951information},
	\begin{equation}
		\sum_{i\in S_{I}}\sum_{j=1}^{n_{i}}\omega_{ij}\alpha_{ij}\log\frac{\alpha_{ij}}{d_{ij}},\label{eq:KL}
	\end{equation}
	subject to the calibration equations (\ref{eq:c1}), (\ref{eq:c2}),
	(\ref{eq:c3}) and (\ref{eq:c4}). The Kullback-Leibler minimum distance
	estimation leads to the exponential tilting estimator (\citealp{kitamura1997information};
	\citealp{imbens1998information}; \citealp{schennach2007point}).
	An advantage of using the exponential tilting estimator is that the
	resulting weights are always non-negative. Also, with exponential
	tilting, the calibration constraints (\ref{eq:c3}) and (\ref{eq:c4})
	can be built into a closed form expression for the weights, and thus
	avoiding solving a large number of equations. See the computation
	below. This reduces the computation burden greatly when there is a
	large number of clusters. By Lagrange Multiplier, the solution to
	(\ref{eq:KL}) is
	\begin{multline*}
		\alpha_{ij}(\lambda_{1},\lambda_{2})=\hat{N}_{i}\frac{A_{ij}d_{ij}\exp(\lambda_{1}X_{ij}A_{ij})}{\sum_{j=1}^{n_{i}}\omega_{ij}A_{ij}d_{ij}\exp(\lambda_{1}X_{ij}A_{ij})}\\
		+\hat{N}_{i}\frac{(1-A_{ij})d_{ij}\exp\{\lambda_{2}X_{ij}(1-A_{ij})\}}{\sum_{j=1}^{n_{i}}\omega_{ij}(1-A_{ij})d_{ij}\exp\{\lambda_{2}X_{ij}(1-A_{ij})\}},
	\end{multline*}
	where $\hat{N}_{i}=\sum_{j=1}^{n_{i}}\omega_{ij}$, and $(\lambda_{1},\lambda_{2})^{T}$
	is the solution to the following equation
	\begin{multline*}
		Q(\lambda_{1},\lambda_{2})=\left(\begin{array}{c}
			Q_{1}(\lambda_{1},\lambda_{2})\\
			Q_{2}(\lambda_{1},\lambda_{2})
		\end{array}\right)\\
		=\left(\begin{array}{c}
			\sum_{i\in S_{I}}\sum_{j=1}^{n_{i}}\omega_{ij}\left\{ A_{ij}\alpha_{ij}(\lambda_{1},\lambda_{2})-1\right\} X_{ij}\\
			\sum_{i\in S_{I}}\sum_{j=1}^{n_{i}}\omega_{ij}\left\{ (1-A_{ij})\alpha_{ij}(\lambda_{1},\lambda_{2})-1\right\} X_{ij}
		\end{array}\right)=0.
	\end{multline*}
\end{description}
From the above calibration algorithm, we obtain an estimate for the
propensity score, $\hat{e}(X_{ij},U_{i})=\alpha_{ij}(\hat{\lambda}_{1},\hat{\lambda}_{2})^{-A_{ij}}\{1-\alpha_{ij}(\hat{\lambda}_{1},\hat{\lambda}_{2})\}^{-1+A_{ij}}$.
The proposed estimator for the average treatment effect $\tau$ is
the weighting estimator (\ref{eq:ps estimator}) with the propensity
score estimates $\hat{e}(X_{ij},U_{i})$, denoted by $\hat{\tau}_{\mathrm{cal}}$. 

\begin{remark}\label{Rmk:1}
	
	The logistic linear fixed effect model in Step 0 is only a working
	model, and the proposed estimator $\hat{\tau}_{\mathrm{cal}}$ does
	not require the specification of this working model to be true. \citet{chan2015globally}
	suggested using an initial set of uniform weights, which controls
	the dispersion of final weights and is less likely to obtain extreme
	final weights. Our simulation studies show that the consistency of
	$\hat{\tau}_{\mathrm{cal}}$ is not sensitive to the choice of the
	initial set of weights. Therefore, $\hat{\tau}_{\mathrm{cal}}$ is
	robust to the specification of this working propensity score model.
	
\end{remark} 

\spacingset{1.45} 

\section{Main results}

Theorem \ref{Thm:unbiasedness} establishes the unbiasedness of the
proposed estimator for the average treatment effect $\tau$, which
indicates that $\tau$ is nonparametrically identifiable by $\hat{\tau}_{\mathrm{cal}}$.
The proof is given in the Appendix. 

\begin{theorem}\label{Thm:unbiasedness}Under Assumptions \ref{assump1}\textendash \ref{assump3-indepence},
	the proposed estimator $\hat{\tau}_{\mathrm{cal}}$ is unbiased of
	$\tau$. 
	
\end{theorem} 

\begin{remark}\label{Rmk:2}
	
	The weighting estimators are often not an efficient estimator. In
	a special case, we found that the weighting estimator is efficient.
	Assume that the potential outcome variables follow linear mixed effects
	models, that is, 
	\begin{equation}
		Y_{ij}(a)=X_{ij}\beta_{a}+U_{i}+e_{ij},\ a=0,1,\label{eq:model1}
	\end{equation}
	with unknown parameters $\beta_{a}$, random effects $U_{i}$ that
	have mean zero, and independent errors $e_{ij}$ such that $E(e_{ij}\mid X_{ij},U_{i})=0$.
	The augmented inverse probability of treatment weighting (AIPTW, \citealp{lunceford2004stratification,bang2005doubly})
	estimator of $\tau$ is
	\begin{eqnarray}
		\hat{\tau}_{\mathrm{AIPTW}} & = & \frac{1}{\hat{N}}\sum_{i\in S_{I}}\sum_{j=1}^{n_{i}}\omega_{ij}\left\{ \frac{A_{ij}Y_{ij}}{\hat{e}(X_{ij},U_{i})}-\frac{(1-A_{ij})Y_{ij}}{1-\hat{e}(X_{ij},U_{i})}\right\} \nonumber \\
		&  & -\frac{1}{\hat{N}}\sum_{i\in S_{I}}\sum_{j=1}^{n_{i}}\omega_{ij}\frac{A_{ij}-\hat{e}(X_{ij},U_{i})}{\hat{e}(X_{ij},U_{i})}(X_{ij}\hat{\beta}_{1}+U_{i})\nonumber \\
		&  & -\frac{1}{\hat{N}}\sum_{i\in S_{I}}\sum_{j=1}^{n_{i}}\omega_{ij}\frac{A_{ij}-\hat{e}(X_{ij},U_{i})}{1-\hat{e}(X_{ij},U_{i})}(X_{ij}\hat{\beta}_{0}+U_{i}),\label{eq:dr estimator}
	\end{eqnarray}
	where $\hat{\beta}_{a}$ is a consistent estimator of $\beta_{a}$
	for $a=0,1$. The AIPTW estimator has been shown to be doubly robust
	\citep{robins1995analysis,robins1997toward,lunceford2004stratification,bang2005doubly,kang2007demystifying}
	in the sense that if either the propensity score model \textit{or
	}the outcome regression model for $Y_{ij}(a)$ is correctly specified,
	the estimator is unbiased. Moreover, if both models are correctly
	specified, the AIPTW estimator is efficient in the setting with iid
	random variables. Notice that by the constraints (\ref{eq:c1}), (\ref{eq:c2}),
	(\ref{eq:c3}), and (\ref{eq:c4}), the augmented terms in (\ref{eq:dr estimator})
	disappear and therefore $\hat{\tau}_{\mathrm{cal}}=\hat{\tau}_{\mathrm{AIPTW}}$. 
	
\end{remark} 

To establish large-sample properties of our estimator, we assume our
sequence of finite populations and samples are as described in \citet{isaki1982survey},
such that the population size $N$ increases but the cluster sample
sizes $M_{i}$ may remain small. In such cases, the number of clusters
$K$ increases linearly with $N$. In other cases, $K$ may increase
at a slower rate than $N$ does. Assume that the sufficient conditions
for the asymptotic normality of the Horvitz-Thompson estimator hold
for the sequence of finite populations and the samples, see \citet{fuller2009sampling}.
For the sequence of designs, we require that the first-order inclusion
probabilities $\pi_{i}\pi_{j|i}$ satisfy $0<\underline{\pi}<\pi_{i}\pi_{j|i}<\bar{\pi}<1$
for some values $\underline{\pi}$ and $\bar{\pi}$, which prevents
producing extremely large design weights that dominate the analyses. 

\begin{theorem}\label{Thm:Asymp Normality}
	
	Under Assumptions \ref{assump1}\textendash \ref{assump4-moment}
	and the above regularity conditions on the sequence of populations,
	samples and designs, the calibrated propensity score weighting estimator
	in (\ref{eq:ps estimator}), subject to constraints (\ref{eq:c1}),
	(\ref{eq:c2}), (\ref{eq:c3}), and (\ref{eq:c4}), satisfies 
	\[
	n^{1/2}N^{-1}(\hat{\tau}_{\mathrm{cal}}-\tau)\rightarrow\mathcal{N}(0,V),
	\]
	as $n\rightarrow\infty$, where 
	\[
	V=nN^{-2}\mathrm{var}\left(\sum_{i\in S_{I}}\sum_{j=1}^{n_{i}}\omega_{ij}\Phi_{ij}\right),
	\]
	with $\Phi_{ij}=\{\alpha_{ij}(\lambda_{1}^{*},\lambda_{2}^{*})A_{ij}(Y_{ij}-B_{1}^{T}X_{ij})+B_{1}^{T}X_{ij}\}-\{\alpha_{ij}(\lambda_{1}^{*},\lambda_{2}^{*})(1-A_{ij})(Y_{ij}-B_{2}^{T}X_{ij})+B_{2}^{T}X_{ij}\}$,
	\begin{eqnarray*}
		B_{1} & = & E\left\{ \sum_{i=1}^{M}\sum_{j=1}^{N_{i}}\alpha_{ij}(\lambda_{1}^{*},\lambda_{2}^{*})\left\{ 1-\frac{\alpha_{ij}(\lambda_{1}^{*},\lambda_{2}^{*})}{n_{i}}\right\} A_{ij}Y_{ij}X_{ij}^{T}\right\} \\
		&  & \times E\left\{ \sum_{i=1}^{M}\sum_{j=1}^{N_{i}}\alpha_{ij}(\lambda_{1}^{*},\lambda_{2}^{*})\left\{ 1-\frac{\alpha_{ij}(\lambda_{1}^{*},\lambda_{2}^{*})}{n_{i}}\right\} A_{ij}X_{ij}X_{ij}^{T}\right\} ^{-1},\\
		B_{2} & = & E\left\{ \sum_{i=1}^{M}\sum_{j=1}^{N_{i}}\alpha_{ij}(\lambda_{1}^{*},\lambda_{2}^{*})\left\{ 1-\frac{\alpha_{ij}(\lambda_{1}^{*},\lambda_{2}^{*})}{n_{i}}\right\} (1-A_{ij})Y_{ij}X_{ij}^{T}\right\} \\
		&  & \times E\left\{ \sum_{i=1}^{M}\sum_{j=1}^{N_{i}}\alpha_{ij}(\lambda_{1}^{*},\lambda_{2}^{*})\left\{ 1-\frac{\alpha_{ij}(\lambda_{1}^{*},\lambda_{2}^{*})}{n_{i}}\right\} (1-A_{ij})X_{ij}X_{ij}^{T}\right\} ^{-1},
	\end{eqnarray*}
	and $(\lambda_{1}^{*},\lambda_{2}^{*})^{T}$ satisfies $E\{Q(\lambda_{1}^{*},\lambda_{2}^{*})\}=0$. 
	
\end{theorem} 

See Appendix for the proof. We now discuss variance estimation. Let
$\phi_{ij}=\alpha_{ij}(\hat{\lambda}_{1},\hat{\lambda}_{2})\{A_{ij}(Y_{ij}-\hat{B}_{1}^{T}X_{ij})-(1-A_{ij})(Y_{ij}-\hat{B}_{2}^{T}X_{ij})\}+(\hat{B}_{1}-\hat{B}_{2})^{T}X_{ij}$,
where 
\begin{eqnarray*}
	\hat{B}_{1} & = & \sum_{i\in S_{I}}\sum_{j=1}^{n_{i}}\omega_{ij}\alpha_{ij}(\hat{\lambda}_{1},\hat{\lambda}_{2})\left\{ 1-\frac{\alpha_{ij}(\hat{\lambda}_{1},\hat{\lambda}_{2})}{n_{i}}\right\} A_{ij}Y_{ij}X_{ij}^{T}\\
	&  & \times\sum_{i\in S_{I}}\sum_{j=1}^{n_{i}}\omega_{ij}\alpha_{ij}(\hat{\lambda}_{1},\hat{\lambda}_{2})\left\{ 1-\frac{\alpha_{ij}(\hat{\lambda}_{1},\hat{\lambda}_{2})}{n_{i}}\right\} A_{ij}X_{ij}X_{ij}^{T},\\
	\hat{B}_{2} & = & \sum_{i\in S_{I}}\sum_{j=1}^{n_{i}}\omega_{ij}\alpha_{ij}(\hat{\lambda}_{1},\hat{\lambda}_{2})\left\{ 1-\frac{\alpha_{ij}(\hat{\lambda}_{1},\hat{\lambda}_{2})}{n_{i}}\right\} (1-A_{ij})Y_{ij}X_{ij}^{T}\\
	&  & \times\sum_{i\in S_{I}}\sum_{j=1}^{n_{i}}\omega_{ij}\alpha_{ij}(\hat{\lambda}_{1},\hat{\lambda}_{2})\left\{ 1-\frac{\alpha_{ij}(\hat{\lambda}_{1},\hat{\lambda}_{2})}{n_{i}}\right\} (1-A_{ij})X_{ij}X_{ij}^{T}.
\end{eqnarray*}
Let $\hat{\tau}_{i}=\sum_{j=1}^{n_{i}}\pi_{j|i}^{-1}\phi_{ij}$ and
\[
\hat{V}_{i}=\sum_{k=1}^{n_{i}}\sum_{l=1}^{n_{i}}\frac{\pi_{kl|i}-\pi_{k|i}\pi_{l|i}}{\pi_{kl|i}}\frac{\phi_{ik}}{\pi_{k|i}}\frac{\phi_{il}}{\pi_{l|i}}.
\]
The variance estimator is
\[
\hat{V}(\hat{\tau}_{\mathrm{cal}})=\frac{1}{\hat{N}^{2}}\left(\sum_{i\in S_{I}}\sum_{j\in S_{I}}\frac{\pi_{ij}-\pi_{i}\pi_{j}}{\pi_{ij}}\frac{\hat{\tau}_{i}}{\pi_{i}}\frac{\hat{\tau}_{j}}{\pi_{j}}+\sum_{i\in S_{I}}\frac{\hat{V}_{i}}{\pi_{i}}\right),
\]
which is design-consistent for $\mathrm{var}(\hat{\tau}_{\mathrm{cal}})$.
The above variance estimation uses linearization. Alternatively, we
can develop replication methods such as Jackknife variance estimation.

\spacingset{1.45} 
\section{Extension to multiple treatments}

There is more and more attention nowadays to the setting with more
than two treatments, which is important and common in empirical practice.
See for example, \citet{imbens2000role}; \citet{robins2000marginal};
\citet{lechner2001identification}; \citet{foster2003propensity};
\citet{hirano2004propensity}; \citet{imai2012causal}; \citet{cole2009consistency};
\citet{cadarette2010confounder}; \citet{cattaneo2010efficient};
\citet{mccaffrey2013inverse}; \citet{rassen2013matching}; and \citet{yang2016propensity}. 

We now extend the potential outcome set up to the case with more than
two treatments as in \citet{imbens2000role}; \citet{lechner2001identification};
\citet{imai2012causal}; and \citet{cattaneo2010efficient}. The treatment
is denoted by $A\in\mathbb{A}=\{1,\ldots T\}$. For each unit $i$
there are $T$ potential outcomes, one for each treatment level, denoted
by $Y_{ij}(a)$, for $a\in\mathbb{A}$. The observed outcome for unit
$i$ is the potential outcome corresponding to the treatment received,
$Y_{ij}=Y_{ij}(A_{ij})$. For the comparison between treatments $a$
and $a'$, the average effect is $\tau(a,a')=E\{Y_{ij}(a)-Y_{ij}(a')\}$.
In this setting, we modify Assumption \ref{assump1} to the following
assumption. 

\begin{assumption}[Ignorability]\label{assump1`}For $a\in\mathbb{A}$,
	$Y_{ij}(a)\bot A_{ij}\mid X_{ij},U_{i}$.
	
\end{assumption} 

Here, we generalize the propensity score to to the multiple treatments
case, following \citet{imbens2000role}: 

\begin{definition}[Generalized Propensity Score] 
	
	The generalized propensity score is the conditional probability of
	receiving each treatment level: $g_{a}(X_{ij},U_{i})=\mathrm{pr}(A=a\mid X_{ij},U_{i})$. 
	
\end{definition} 

The overlap assumption is modified as follows.

\begin{assumption}[Overlap]\label{assump3`}For $a\in\mathbb{A}$,
	$g_{a}(X_{ij},U_{i})>\underline{e}>0$. 
	
\end{assumption} 

We consider a parametric working model for the generalized propensity
score, and calibrate the generalized propensity score to satisfy certain
constraints. Since we have 
\[
E\left\{ \frac{I(A=a)}{g_{a}(X,U)}X\right\} =E(X),\ E\left\{ \frac{I(A=a)}{g_{a}(X,U)}U\right\} =E(U),\ a\in\mathbb{A}.
\]

Based on the sample, for the estimated generalized propensity score
$\hat{g}_{a}(X_{ij},U_{i})$, we would impose the following constraints,
\begin{equation}
	\sum_{i\in S_{I}}\sum_{j=1}^{n_{i}}\omega_{ij}\frac{I(A_{ij}=a)}{\hat{g}_{a}(X_{ij},U_{i})}X_{ij}=\sum_{i\in S_{I}}\sum_{j=1}^{n_{i}}\omega_{ij}X_{ij},\label{eq:c1-m}
\end{equation}
\begin{equation}
	\sum_{j=1}^{n_{i}}\omega_{ij}\frac{I(A_{ij}=a)}{\hat{g}_{a}(X_{ij},U_{i})}=\sum_{j=1}^{n_{i}}\omega_{ij},\ (i=1,\ldots,K,a\in\mathbb{A}).\label{eq:c2-m}
\end{equation}

We now discuss the specific steps for computation. 
\begin{description}
	\item [{Step0.}] Consider a working model, for example a fixed effect multinomial
	logistic regression model with a cluster-level main effect, fitted
	to $(A_{ij},X_{ij},\delta_{i})$ where $\delta_{i}$ is the cluster
	indicator. We obtain an initial estimate for the generalized propensity
	score, 
	\[
	g_{a}^{0}(X_{ij},U_{i})=\frac{\exp(X_{ij}\hat{\beta}_{a}+\hat{\gamma}_{i})}{\sum_{a=1}^{T}\exp(X_{ij}\hat{\beta}_{a}+\hat{\gamma}_{i})},\ a\in\mathbb{A},
	\]
	where $(\hat{\beta}_{1},\ldots\hat{\beta}_{T})$ and $(\hat{\gamma}_{1},\ldots,\hat{\gamma}_{m})$
	are the fitted estimates. This provides an initial set of inverse
	propensity score weights $\mathbb{W}^{0}=\{d_{ij};i\in A_{I},j=1,\ldots,n_{i}\}$,
	with $d_{ij}=1/g_{A_{ij}}^{0}(X_{ij},U_{i})$.
	\item [{Step1.}] We modify the initial set of weights $\mathbb{W}^{0}$
	to a new set of weights $\mathbb{W}=\{\alpha_{ij};i\in S_{I},j=1,\ldots,n_{i}\}$,
	by minimizing the Kullback-Leibler distance, 
	\[
	\sum_{i\in S_{I}}\sum_{j=1}^{n_{i}}\omega_{ij}\alpha_{ij}\log\frac{\alpha_{ij}}{d_{ij}},
	\]
	subject to the calibration equations (\ref{eq:c1-m}) and (\ref{eq:c2-m}).
	By Lagrange Multiplier, the solution is
	\[
	\alpha_{ij}(\lambda)=\hat{N}_{i}\sum_{a=1}^{T}\frac{I(A_{ij}=a)d_{ij}\exp(\lambda_{a}^{T}X_{ij}A_{ij})}{\sum_{j=1}^{n_{i}}\omega_{ij}I(A_{ij}=a)d_{ij}\exp(\lambda_{a}^{T}X_{ij}A_{ij})},
	\]
	where $\lambda=(\lambda_{1},\ldots,\lambda_{T})^{T}$ is the solution
	to the following equation
	\begin{equation}
		Q(\lambda)=\left(\begin{array}{c}
			Q_{1}(\lambda)\\
			\vdots\\
			Q_{T}(\lambda)
		\end{array}\right)=\left(\begin{array}{c}
		\sum_{i\in S_{I}}\sum_{j=1}^{n_{i}}\omega_{ij}\{I(A_{ij}=1)\alpha_{ij}(\lambda)-1\}X_{ij}\\
		\vdots\\
		\sum_{i\in S_{I}}\sum_{j=1}^{n_{i}}\omega_{ij}\{I(A_{ij}=T)\alpha_{ij}(\lambda)-1\}X_{ij}
	\end{array}\right)=0.\label{eq:Q-m}
\end{equation}
\end{description}
We obtain a final estimate for the generalized propensity score, $\hat{g}_{A_{ij}}(X_{ij},U_{i})=1/\alpha_{ij}(\hat{\lambda})$.
The calibrated weighting estimator for $\tau(a,a')$ is
\begin{equation}
	\hat{\tau}_{\mathrm{cal}}(a,a')=\sum_{i\in S_{I}}\sum_{j=1}^{n_{i}}\omega_{ij}\left\{ \frac{I(A_{ij}=a)}{\hat{g}_{a}(X_{ij},U_{i})}Y_{ij}-\frac{I(A_{ij}=a')}{\hat{g}_{a'}(X_{ij},U_{i})}Y_{ij}\right\} .\label{eq:cal-m}
\end{equation}
\begin{theorem}\label{Thm-m}Under Assumptions \ref{assump3-indepence}\textendash \ref{assump3`},
	and the same regularity conditions as in Theorem \ref{Thm:Asymp Normality},
	the calibrated propensity score weighting estimator $\hat{\tau}_{\mathrm{cal}}(a,a')$
	in (\ref{eq:cal-m}) subject to constraints (\ref{eq:c1-m}) and (\ref{eq:c2-m})
	is unbiased of $\tau(a,a')$, and satisfies 
	\[
	n^{1/2}N^{-1}\{\hat{\tau}_{\mathrm{cal}}(a,a')-\tau(a,a')\}\rightarrow\mathcal{N}\left(0,V(a,a')\right),
	\]
	as $n\rightarrow\infty$, where 
	\[
	V(a,a')=nN^{-2}\mathrm{var}\left(\sum_{i\in S_{I}}\sum_{j=1}^{n_{i}}\omega_{ij}\Phi_{ij}(a,a')\right),
	\]
	with $\Phi_{ij}(a,a')=\{\alpha_{ij}(\lambda^{*})I(A_{ij}=a)(Y_{ij}-B_{a}^{T}X_{ij})+B_{a}^{T}X_{ij}\}-\{\alpha_{ij}(\lambda^{*})I(A_{ij}=a')(Y_{ij}-B_{a'}^{T}X_{ij})+B_{a'}^{T}X_{ij}\}$,
	\begin{eqnarray*}
		B_{a} & = & E\left\{ \sum_{i=1}^{M}\sum_{j=1}^{N_{i}}\alpha_{ij}(\lambda^{*})\left\{ 1-\frac{\alpha_{ij}(\lambda^{*})}{n_{i}}\right\} I(A_{ij}=a)Y_{ij}X_{ij}^{T}\right\} \\
		&  & \times E\left\{ \sum_{i=1}^{M}\sum_{j=1}^{N_{i}}\alpha_{ij}(\lambda^{*})\left\{ 1-\frac{\alpha_{ij}(\lambda^{*})}{n_{i}}\right\} I(A_{ij}=a)X_{ij}X_{ij}^{T}\right\} ^{-1},
	\end{eqnarray*}
	and $\lambda^{*}$ satisfies $E\{Q(\lambda^{*})\}=0$, with $Q(\lambda)$
	defined in (\ref{eq:Q-m}).
	
\end{theorem}

The proof of Theorem \ref{Thm-m} is similar to that of Theorems \ref{Thm:unbiasedness}
and \ref{Thm:Asymp Normality}, and therefore is omitted to avoid
redundancy. Similarly, the variance estimator of $\hat{\tau}_{\mathrm{cal}}(a,a')$
can be developed accordingly. Let $\phi_{ij}(a,a')=\alpha_{ij}(\hat{\lambda}_{1},\hat{\lambda}_{2})\{I(A_{ij}=a)(Y_{ij}-\hat{B}_{a}^{T}X_{ij})-I(A_{ij}=a')(Y_{ij}-\hat{B}_{a'}^{T}X_{ij})\}+(\hat{B}_{a}-\hat{B}_{a'})^{T}X_{ij}$,
where 
\begin{eqnarray*}
	\hat{B}_{a} & = & \sum_{i\in S_{I}}\sum_{j=1}^{n_{i}}\omega_{ij}\alpha_{ij}(\hat{\lambda}_{1},\hat{\lambda}_{2})\left\{ 1-\frac{\alpha_{ij}(\hat{\lambda}_{1},\hat{\lambda}_{2})}{n_{i}}\right\} I(A_{ij}=a)Y_{ij}X_{ij}^{T}\\
	&  & \times\sum_{i\in S_{I}}\sum_{j=1}^{n_{i}}\omega_{ij}\alpha_{ij}(\hat{\lambda}_{1},\hat{\lambda}_{2})\left\{ 1-\frac{\alpha_{ij}(\hat{\lambda}_{1},\hat{\lambda}_{2})}{n_{i}}\right\} I(A_{ij}=a)X_{ij}X_{ij}^{T}.
\end{eqnarray*}
Let $\hat{\tau}_{i}(a,a')=\sum_{j=1}^{n_{i}}\pi_{j|i}^{-1}\phi_{ij}(a,a')$
and 
\[
\hat{V}_{i}(a,a')=\sum_{k=1}^{n_{i}}\sum_{l=1}^{n_{i}}\frac{\pi_{kl|i}-\pi_{k|i}\pi_{l|i}}{\pi_{kl|i}}\frac{\phi_{ik}(a,a')}{\pi_{k|i}}\frac{\phi_{il}(a,a')}{\pi_{l|i}}.
\]
The variance estimator is
\[
\hat{V}\left\{ \hat{\tau}_{\mathrm{cal}}(a,a')\right\} =\frac{1}{\hat{N}^{2}}\left\{ \sum_{i\in S_{I}}\sum_{j\in S_{I}}\frac{\pi_{ij}-\pi_{i}\pi_{j}}{\pi_{ij}}\frac{\hat{\tau}_{i}(a,a')}{\pi_{i}}\frac{\hat{\tau}_{j}(a,a')}{\pi_{j}}+\sum_{i\in S_{I}}\frac{\hat{V}_{i}(a,a')}{\pi_{i}}\right\} .
\]

\spacingset{1.45} 
\section{Simulation Study }

We conducted two simulation studies to evaluate the finite-sample
performance of the proposed estimator. We first generated finite populations
and then selected a sample from each finite population using a two-stage
cluster sampling design. 

In the first setting, the potential outcomes were generated according
to linear mixed effect models, $Y_{ij}(0)=X_{ij}+U_{i}+e_{ij}$ and
$Y_{ij}(1)=X_{ij}+\tau+\tau U_{i}+e_{ij}$, with $\tau=2$, $U_{i}\sim N(0,1)$,
$X_{ij}\sim N(0,1)$, $e_{ij}\sim N(0,1)$, $U_{i}$, $X_{ij}$, $e_{ij}$
are independent, $i=1,\ldots,M=10,000$, $j=1,\ldots,N_{i}$, and
$N_{i}$ is the integer part of $500\exp(2+U_{i})/\{1+\exp(2+U_{i})\}$.
The population cluster sizes range from $100$ to $500$. The parameter
of interest is $\tau=E\{Y_{ij}(1)-Y_{ij}(0)\}$. We considered three
propensity score models, $\mathrm{pr}(A_{ij}=1\mid X_{ij};U_{i})=h(\gamma_{0}+\gamma_{1}U_{i}+X_{ij})$,
with $h(\cdot)$ being the inverse logit, probit and complementary
log-log link function. The observed outcome is $Y_{ij}=A_{ij}Y_{ij}(1)+(1-A_{ij})Y_{ij}(0)$.
From each realized population, $m$ clusters were sampled by PPS (Probability-Proportional-to-Size)
sampling with the measure of size $N_{i}$. So the first-order inclusion
probability of selecting cluster $i$ is equal to $\pi_{i}=mN_{i}/\sum_{i=1}^{I}N_{i}$,
which implicitly depends on the unobserved random effect. Once the
clusters were sampled, the $n_{i}$ units in the $i$th sampled cluster
were sampled by Poison sampling with the corresponding first-order
inclusion probabilities $\pi_{j|i}=nz_{ij}/(\sum_{j=1}^{M_{i}}z_{ij})$,
where $z_{ij}=0.5$ if $e_{ij}<0$ and 1 if $e_{ij}>0$. With this
sampling design, the units with $e_{ij}>0$ were sampled with a chance
twice as big as the units with $e_{ij}<0$. We considered three combinations
of the number of clusters $m$ and the cluster size $n$: (i) $(m,n)=(50,50)$;
(ii) $(m,n)=(100,30)$, with a large number of small clusters; and
(iii) $(m,n)=(30,100)$, with a small number of large clusters. 

In the second setting, all data-generating mechanisms were the same
with the first setting, except that the potential outcomes were generated
according to logistic linear mixed effect models, $Y_{ij}(0)\sim\mathrm{Bernoulli}(p_{ij}^{0})$
with $\mathrm{logit}(p_{ij}^{0})=X_{ij}+U_{i}$ and $Y_{ij}(1)\sim\mathrm{Bernoulli}(p_{ij}^{1})$
with $\mathrm{logit}(p_{ij}^{1})=X_{ij}+\tau+\tau u_{i}$, and moreover,
in the 2-stage sampling, $\pi_{j|i}=nz_{ij}/(\sum_{j=1}^{M_{i}}z_{ij})$,
where $z_{ij}=0.5$ if $Y_{ij}=0$ and 1 if $Y_{ij}=1$. With this
sampling design, the units with $Y_{ij}=1$ were sampled with a chance
twice as big as the units with $Y_{ij}=0$. 

We computed four estimators for $\tau$: (i) $\hat{\tau}_{\mathrm{simp}}$,
the simple design-weighted estimator without propensity score adjustment;
(ii) $\hat{\tau}_{\mathrm{fix}}$, the weighting estimator (\ref{eq:ps estimator})
with the propensity score estimated by a logistic linear fixed effect
model with a cluster-level main effect; (iii) $\hat{\tau}_{\mathrm{ran}}$,
the weighting estimator (\ref{eq:ps estimator}) with the propensity
score estimated by a logistic linear mixed effect model where the
cluster effect is random; and (iv) $\hat{\tau}_{\mathrm{cal}}$, the
proposed estimator with calibrations. We reported empirical biases,
variances, coverages for $95\%$ confidence intervals from $1,000$
simulated datasets. 

Table \ref{tab:1} shows the simulation results. The simple estimator
shows large biases across difference scenarios, even adjusted for
sampling design. This suggests that covariate distributions are different
between treatment groups in the finite population, contributing to
the bias. $\hat{\tau}_{\mathrm{fix}}$ works well under Scenario 1
with the linear mixed effect model for the outcome and the logistic
linear mixed effect model for the propensity score; however, its performance
is not satisfactory under other scenarios. This is because except
for Scenario 1, weighting by the logistic linear fixed effect model
does not balance the covariate distributions in the finite population.
Moreover, $\hat{\tau}_{\mathrm{fix}}$ shows largest variance among
the four estimators. This is because for a moderate or large number
of clusters, there are many free parameters and the propensity score
estimates may not be stable. For $\hat{\tau}_{\mathrm{ran}}$, we
assume that the cluster effect is random, which reduces the number
of free parameters greatly. As a result, $\hat{\tau}_{\mathrm{ran}}$
shows less variability than $\hat{\tau}_{\mathrm{fix}}$. Nonetheless,
both $\hat{\tau}_{\mathrm{fix}}$ and $\hat{\tau}_{\mathrm{ran}}$
can not control the bias well. The proposed calibrated propensity
score weighting estimator is essentially unbiased of $\tau$ under
all scenarios, and the empirical coverages are close to the nominal
coverage. Here, we used a working model, a logistic linear fixed effect
model, to provide an initial set of weights. But the consistency of
the estimator does not rely on this working model. When the true propensity
score is probit or complementary log-log model, $\hat{\tau}_{\mathrm{cal}}$
is still consistent, confirming our theoretical results. We also examined
an initial set of uniform weights and did not find results that were
meaningfully different from those reported above.

\spacingset{1.45} 
\section{An Application}

We examined the 2007\textendash 2010 BMI surveillance data from Pennsylvania
Department of Health to investigate the effect of School Body Mass
Index Screening (SBMIS) on the annual overweight and obesity prevalence
in elementary schools in Pennsylvania. Early studies have shown that
SBMIS has been associated with increased parental awareness of child
weight \citep{harris2009effect,ebbeling2012randomized}. However,
there have been mixed findings about effects of screening on reducing
prevalence of overweight and obesity \citep{harris2009effect,thompson2009arkansas}. 

The data includes $493$ school districts in Pennsylvania. The baseline
is the school year 2007. The schools are clustered by two factors:
location (rural, suburban, and urban), and population density (low,
median, and high). This results in five clusters: rural-low, rural-median,
rural-high, suburban-high, and urban-high. Let $A=1$ if the school
implemented SBMIS, and $A=0$ if the school did not. In this dataset,
$63\%$ of schools implemented SBMIS, and the percentages of schools
implemented SBMIS across the clusters are from $45\%$ to $70\%$,
indicating cluster-level heterogeneity of treatment. The outcome variable
$Y$ is the annual overweight and obesity prevalence for each district
by dividing the number with Body Mass Index (BMI) $>85$th by the
total number of students screened for each district in the school
year 2010. For each school, we obtain individual characteristics including
the baseline prevalence of overweight and obesity $X_{1}$, and percentage
of reduced and free lunch $X_{2}$. 

For a direct comparison, the average difference of the prevalence
of overweight and obesity for schools that implemented SBMIS and those
that did not is $8.78\%$. This unadjusted difference in the prevalence
of overweight and obesity ignores differences in individual covariates
and cluster characteristics. \textcolor{black}{Standard propensity
	score analyses including $X_{1}$ and $X_{2}$ would account for these
	observed differences, but there may also be unobserved cluster-level
	confounders. When such unmeasured confounders exist but are omitted
	from the propensity score model, the ensuing analysis will fail to
	control for the bias. }

We consider the propensity score models that also account for cluster
effects. Specifically, we consider three methods: (i) a logistic linear
fixed effect model with linear predictors including $X_{1}$, $X_{2}$,
and a fixed intercept for each cluster; (ii) a logistic linear mixed
effect model with linear predictors including fixed effects $X_{1}$,
$X_{2}$, and a random effect for each cluster; (iii) the proposed
calibrated propensity score. Using the estimated propensity score,
we estimate $\tau=E\{Y(1)-Y(0)\}$ by the weighting method. 

Table \ref{tab:Results-3} displays the standardized differences of
means for covariates $X_{1}$ and $X_{2}$ between the treated and
the control for each cluster and the whole population, standardized
by the standard errors in the whole population. Without any adjustment,
there are large differences in means for $X_{1}$ and $X_{2}$. All
three propensity score weighting methods improve the balances for
$X_{1}$ and $X_{2}$. For this specific dataset, the three methods
for modeling and estimating the propensity score are similar in balancing
the covariate distributions between the treated and the control. Table
\ref{tab:Results-2} displays point estimates and variance estimates
based on $500$ bootstrap replicates. The simple estimator shows that
the screening has significant effect in reducing the prevalence of
overweight and obesity. However, this may be due to the observed confounders
and the unobserved cluster-level confounders. After adjusting for
the confounders, the screening does not have significant effect. 

\spacingset{1.45} 
\section{Discussion}

Inverse probability of treatment weighting (IPTW) estimator is not
efficient in general. Semiparametric efficiency bounds for estimating
the average treatment effects in the setting with iid random variables
were derived by \citet{hahn1998role}. He showed that the efficient
influence function for the average treatment effect depends on both
the propensity score and the outcome model. An important implication
is that combining the propensity score model and the outcome regression
model can improve efficiency of the IPTW estimator. For clustered
data, since the data are correlated through the random cluster effects,
the efficiency theory established for the iid data is not applicable.
Developing semiparametric efficiency theory for clustered data is
interesting. This extension will be a subject of future work. 

In this article, we assumed that there is no interference between
units. This setup is not uncommon. In our application, the treatment
was implemented school-wise. The potential outcomes for one school
are likely to be unaffected by the treatments implemented at other
schools, and therefore the assumption of no interference is likely
to hold. However, in other settings this assumption may not hold.
A classical example is given in infectious diseases \citep{ross1916application,hudgens2008toward},
where whether one person becomes infected depends on who else in the
population is vaccinated. Extension of our calibration estimation
to take the interference structure into account in these settings
is also an interesting topic for future research. 

In addition to propensity score weighting, propensity score has been
used for subclassification \citep{rosenbaum1984reducing,rosenbaum1991characterization}
and matching \citep{rosenbaum1985constructing,abadie2006large}. In
the causal inference and missing data literature, previous simulations
have found that weighting estimators can have high variability, see
for example, \citet{foster2003propensity} and \citet{frolich2004finite}
found that the weighting estimator was inferior to matching estimators
in terms of mean squared error. Therefore, developing subclassification
and matching estimator for clustered data is important, which will
be another topic for future research. 

\spacingset{1.45} 
\section*{Acknowledgments }

We would like to thank Peng Ding for insightful and fruitful discussions.






\appendix

\section*{Appendix}

\section*{Appendix A. Proof of Theorem \ref{Thm:unbiasedness}}

\begin{proof} 
	
Write 
\begin{eqnarray}
	E(\text{\ensuremath{\hat{\tau}_{\mathrm{cal}}}}) & = & E\left[\frac{1}{\hat{N}}\sum_{i\in S_{I}}\sum_{j=1}^{n_{i}}\omega_{ij}\left\{ \frac{A_{ij}Y_{ij}}{\hat{e}(X_{ij},U_{i})}-\frac{(1-A_{ij})Y_{ij}}{1-\hat{e}(X_{ij},U_{i})}\right\} \right]\nonumber \\
	& = & E\left[\frac{1}{\hat{N}}\sum_{i\in S_{I}}\sum_{j=1}^{n_{i}}\omega_{ij}\left\{ \frac{A_{ij}Y_{ij}}{\hat{e}(X_{ij},U_{i})}-Y_{ij}(1)\right\} \right]\nonumber \\
	&  & -E\left[\frac{1}{\hat{N}}\sum_{i\in S_{I}}\sum_{j=1}^{n_{i}}\omega_{ij}\left\{ \frac{(1-A_{ij})Y_{ij}}{1-\hat{e}(X_{ij},U_{i})}-Y_{ij}(0)\right\} \right]\nonumber \\
	&  & +E\left[\frac{1}{\hat{N}}\sum_{i\in S_{I}}\sum_{j=1}^{n_{i}}\omega_{ij}\{Y_{ij}(1)-Y_{ij}(0)\}\right]\nonumber \\
	& \cong & E\left[\frac{1}{\hat{N}}\sum_{i\in S_{I}}\sum_{j=1}^{n_{i}}\omega_{ij}\left\{ \frac{A_{ij}}{\hat{e}(X_{ij},U_{i})}-1\right\} Y_{ij}(1)\right]\nonumber \\
	&  & -E\left[\frac{1}{\hat{N}}\sum_{i\in S_{I}}\sum_{j=1}^{n_{i}}\omega_{ij}\left\{ \frac{(1-A_{ij})}{1-\hat{e}(X_{ij},U_{i})}-1\right\} Y_{ij}(0)\right]+\tau,\label{eq:s1}
\end{eqnarray}
where $B\cong C$ means that $B=C+o_{p}(1)$ for random variables
and $B=C+o(1)$ for non-random variables, and the approximation in
(\ref{eq:s1}) follows from the consistency assumption and that $\hat{N}^{-1}\sum_{i\in S_{I}}\sum_{j=1}^{n_{i}}\omega_{ij}\{Y_{ij}(1)-Y_{ij}(0)\}$
is design-model consistent for $\tau$. Therefore, to show that $\hat{\tau}_{\mathrm{cal}}$
is unbiased for$\tau$, it is sufficient to show that 
\begin{eqnarray*}
	E\left[\frac{1}{\hat{N}}\sum_{i\in S_{I}}\sum_{j=1}^{n_{i}}\omega_{ij}\left\{ \frac{A_{ij}}{\hat{e}(X_{ij},U_{i})}-1\right\} Y_{ij}(1)\right] & = & 0,\\
	E\left[\frac{1}{\hat{N}}\sum_{i\in S_{I}}\sum_{j=1}^{n_{i}}\omega_{ij}\left\{ \frac{(1-A_{ij})}{1-\hat{e}(X_{ij},U_{i})}-1\right\} Y_{ij}(0)\right] & = & 0.
\end{eqnarray*}
Since by the calibration equations (\ref{eq:c3}) and (\ref{eq:c4}),
for any functions $\mu_{0}(U_{i})$ and $\mu_{1}(U_{i})$, we have
\begin{eqnarray}
	\frac{1}{\hat{N}}\sum_{i\in S_{I}}\sum_{j=1}^{n_{i}}\omega_{ij}\left\{ \frac{A_{ij}}{\hat{e}(X_{ij},U_{i})}-1\right\} \mu_{1}(U_{i}) & = & 0,\label{eq:s2}\\
	\frac{1}{\hat{N}}\sum_{i\in S_{I}}\sum_{j=1}^{n_{i}}\omega_{ij}\left\{ \frac{(1-A_{ij})}{1-\hat{e}(X_{ij},U_{i})}-1\right\} \mu_{0}(U_{i}) & = & 0.\label{eq:s3}
\end{eqnarray}
We shall rely on the above equations to show the unbiasedness of $\hat{\tau}_{\mathrm{cal}}$.
Define for $a=0,1$, 
\begin{eqnarray}
	\mu_{a}(U_{i}) & =\frac{\int q_{a}(x,U_{i})E\left\{ Y_{ij}(a)\mid x,U_{i}\right\} f(x)dx}{\int q_{a}(x,U_{i})f(x)dx} & ,\label{eq:nu(a)}
\end{eqnarray}
where $f(x)$ is the density of $X$, 
\[
q_{1}(X_{ij},U_{i})=E\left\{ \frac{A_{ij}}{\hat{e}(X_{ij},U_{i})}-1\mid X_{ij},U_{i}\right\} ,
\]
and
\[
q_{0}(X_{ij},U_{i})=E\left\{ \frac{1-A_{ij}}{1-\hat{e}(X_{ij},U_{i})}-1\mid X_{ij},U_{i}\right\} .
\]
Now, following (\ref{eq:s1}), 
\begin{eqnarray}
	E(\hat{\tau}_{\mathrm{cal}})-\tau & = & E\left[\frac{1}{\hat{N}}\sum_{i\in S_{I}}\sum_{j=1}^{n_{i}}\omega_{ij}\left\{ \frac{A_{ij}}{\hat{e}(X_{ij},U_{i})}-1\right\} Y_{ij}(1)\right]\nonumber \\
	&  & -E\left[\frac{1}{\hat{N}}\sum_{i\in S_{I}}\sum_{j=1}^{n_{i}}\omega_{ij}\left\{ \frac{(1-A_{ij})}{1-\hat{e}(X_{ij},U_{i})}-1\right\} Y_{ij}(0)\right]\nonumber \\
	&  & E\left[\frac{1}{\hat{N}}\sum_{i\in S_{I}}\sum_{j=1}^{n_{i}}\omega_{ij}\left\{ \frac{A_{ij}}{\hat{e}(X_{ij},U_{i})}-1\right\} \mbox{\ensuremath{\mu}}_{1}(U_{i})\right]\nonumber \\
	&  & +E\left[\frac{1}{\hat{N}}\sum_{i\in S_{I}}\sum_{j=1}^{n_{i}}\omega_{ij}\left\{ \frac{1-A_{ij}}{1-\hat{e}(X_{ij},U_{i})}-1\right\} \mbox{\ensuremath{\mu}}_{0}(U_{i})\right]\nonumber \\
	&  & +E\left[\frac{1}{\hat{N}}\sum_{i\in S_{I}}\sum_{j=1}^{n_{i}}\omega_{ij}\left\{ \frac{A_{ij}}{\hat{e}(X_{ij},U_{i})}-1\right\} \left\{ Y_{ij}(1)-\mbox{\ensuremath{\mu}}_{1}(U_{i})\right\} \right]\nonumber \\
	&  & +E\left[\frac{1}{\hat{N}}\sum_{i\in S_{I}}\sum_{j=1}^{n_{i}}\omega_{ij}\left\{ \frac{1-A_{ij}}{1-\hat{e}(X_{ij},U_{i})}-1\right\} \left\{ Y_{ij}(0)-\mbox{\ensuremath{\mu}}_{0}(U_{i})\right\} \right]\nonumber \\
	& = & E\left[\frac{1}{\hat{N}}\sum_{i\in S_{I}}\sum_{j=1}^{n_{i}}\omega_{ij}\left\{ \frac{A_{ij}}{\hat{e}(X_{ij},U_{i})}-1\right\} \left\{ Y_{ij}(1)-\mbox{\ensuremath{\mu}}_{1}(U_{i})\right\} \right]\nonumber \\
	&  & +E\left[\frac{1}{\hat{N}}\sum_{i\in S_{I}}\sum_{j=1}^{n_{i}}\omega_{ij}\left\{ \frac{1-A_{ij}}{1-\hat{e}(X_{ij},U_{i})}-1\right\} \left\{ Y_{ij}(0)-\mbox{\ensuremath{\mu}}_{0}(U_{i})\right\} \right],\label{eq:unbiasedness}
\end{eqnarray}
where the second equality follows (\ref{eq:s2}) and (\ref{eq:s3}).
Now we first consider the first term in in (\ref{eq:unbiasedness}),
\begin{eqnarray}
	&  & E\left[\frac{1}{\hat{N}}\sum_{i\in S_{I}}\sum_{j=1}^{n_{i}}\omega_{ij}\left\{ \frac{A_{ij}}{\hat{e}(X_{ij},U_{i})}-1\right\} \left\{ Y_{ij}(1)-\mbox{\ensuremath{\mu}}_{1}(U_{i})\right\} \right]\nonumber \\
	& \cong & E\left[\frac{1}{N}\sum_{i=1}^{M}\sum_{j=1}^{N_{i}}\left\{ \frac{A_{ij}}{\hat{e}(X_{ij},U_{i})}-1\right\} \left\{ Y_{ij}(1)-\mbox{\ensuremath{\mu}}_{1}(U_{i})\right\} \right]\nonumber \\
	& = & E\left(\frac{1}{N}\sum_{i=1}^{M}\sum_{j=1}^{N_{i}}\left\{ \frac{A_{ij}}{\hat{e}(X_{ij},U_{i})}-1\right\} \left[E\left\{ Y_{ij}(1)\mid X_{ij},U_{i}\right\} -\mbox{\ensuremath{\mu}}_{1}(U_{i})\right]\right)\nonumber \\
	& = & E\left(\frac{1}{N}\sum_{i=1}^{M}\sum_{j=1}^{N_{i}}E\left\{ \frac{A_{ij}}{\hat{e}(X_{ij},U_{i})}-1\mid X_{ij},U_{i}\right\} \left[E\left\{ Y_{ij}(1)\mid X_{ij},U_{i}\right\} -\mbox{\ensuremath{\mu}}_{1}(U_{i})\right]\right)\nonumber \\
	& = & E\left(\frac{1}{N}\sum_{i=1}^{M}\sum_{j=1}^{N_{i}}q_{1}(X_{ij},U_{i})\left[E\left\{ Y_{ij}(1)\mid X_{ij},U_{i}\right\} -\mbox{\ensuremath{\mu}}_{1}(U_{i})\right]\right)\nonumber \\
	& = & 0,\label{eq:part1}
\end{eqnarray}
where the second equality follows from Assumption \ref{assump1} (note
that $\hat{e}(X_{ij},U_{i})$ does not rely on the outcome variable),
and the last equality follows from Assumption \ref{assump3-indepence}
and the definition of $\mu_{a}(U_{i})$ in (\ref{eq:nu(a)}). Similarly,
we can show that the second term in (\ref{eq:unbiasedness}) is zero.
Combining the above results with (\ref{eq:unbiasedness}), we obtain
that $E(\hat{\tau}_{\mathrm{cal}}-\hat{\tau})=0$, leading to $E(\hat{\tau}_{\mathrm{cal}})=\tau$.

\end{proof} 

\section*{Appendix B. Proof of Theorem 2}

\begin{proof} 
	
	Let $Q(\hat{\lambda}_{1},\hat{\lambda}_{2})=0$, and $(\lambda_{1}^{*},\lambda_{2}^{*})$
	satisfy $E\{Q(\lambda_{1}^{*},\lambda_{2}^{*})\}=0$. By linearization,
	we obtain 
	\begin{eqnarray*}
		\hat{\tau}_{\mathrm{cal}} & = & \hat{\tau}_{\mathrm{cal}}(\hat{\lambda}_{1},\hat{\lambda}_{2})\cong\hat{\tau}_{\mathrm{cal}}(\lambda_{1}^{*},\lambda_{2}^{*})-E\left\{ \frac{\partial\hat{\tau}_{\mathrm{cal}}(\lambda_{1}^{*},\lambda_{2}^{*})}{\partial(\lambda_{1},\lambda_{2})^{T}}\right\} \left\{ \frac{\partial Q(\lambda_{1}^{*},\lambda_{2}^{*})}{\partial(\lambda_{1},\lambda_{2})^{T}}\right\} ^{-1}Q(\lambda_{1}^{*},\lambda_{2}^{*})\\
		& = & \hat{\tau}_{\mathrm{cal}}(\lambda_{1}^{*},\lambda_{2}^{*})-B_{1}^{T}Q_{1}(\lambda_{1}^{*},\lambda_{2}^{*})-B_{2}^{T}Q_{2}(\lambda_{1}^{*},\lambda_{2}^{*})\\
		& \cong & \frac{1}{N}\sum_{i\in A_{I}}\sum_{j=1}^{n_{i}}\omega_{ij}\left\{ \alpha_{ij}(\lambda_{1}^{*},\lambda_{2}^{*})A_{ij}(Y_{ij}-B_{1}^{T}X_{ij})+B_{1}^{T}X_{ij}\right\} \\
		&  & -\frac{1}{N}\sum_{i\in A_{I}}\sum_{j=1}^{n_{i}}\omega_{ij}\left\{ \alpha_{ij}(\lambda_{1}^{*},\lambda_{2}^{*})(1-A_{ij})(Y_{ij}-B_{2}^{T}X_{ij})+B_{2}^{T}X_{ij}\right\} ,\\
		& = & \frac{1}{N}\sum_{i\in A_{I}}\sum_{j=1}^{n_{i}}\omega_{ij}\Phi_{ij}.
	\end{eqnarray*}
\end{proof} 

\bibliographystyle{dcu}
\bibliography{ci}

\newpage

\spacingset{1}
\begin{table}
	\caption{\label{tab:1}Simulation results: bias, variance (var$/10^{3}$) and
		coverage (cvg$/100$) of $95\%$ confidence intervals based on $1,000$
		Monte Carlo samples; the outcome is linear and logistic linear mixed
		effect model and the propensity score is logistic, probit or complementary
		log-log (C-loglog).}
	\centering{}%
	\begin{tabular}{cccccccccccc}
		\hline 
		& \multicolumn{3}{c}{{\footnotesize{}$(m,n)=(50,50)$}} &  & \multicolumn{3}{c}{{\footnotesize{}$(m,n)=(100,30)$}} &  & \multicolumn{3}{c}{{\footnotesize{}$(m,n)=(30,100)$}}\tabularnewline
		{\footnotesize{}Method} & {\footnotesize{}bias} & {\footnotesize{}var} & {\footnotesize{}cvg} &  & {\footnotesize{}bias} & {\footnotesize{}var} & {\footnotesize{}cvg} &  & {\footnotesize{}bias} & {\footnotesize{}var} & {\footnotesize{}cvg}\tabularnewline
		\hline 
		\multicolumn{12}{c}{{\footnotesize{}Scenario 1: Linear outcome \& Logistic propensity
				score}}\tabularnewline
		{\footnotesize{}$\hat{\tau}_{\mathrm{simp}}$} & \textcolor{black}{\footnotesize{}-0.37} & {\footnotesize{}22} & {\footnotesize{}27.4 } &  & {\footnotesize{}-0.38} & {\footnotesize{}12} & {\footnotesize{}8.7} &  & {\footnotesize{}-0.38} & {\footnotesize{}35} & {\footnotesize{}42.3}\tabularnewline
		{\footnotesize{}$\hat{\tau}_{\mathrm{fix}}$} & {\footnotesize{}-0.01} & \textcolor{black}{\footnotesize{}36} & \textcolor{black}{\footnotesize{}95.6} &  & \textcolor{black}{\footnotesize{}0.00} & \textcolor{black}{\footnotesize{}21} & \textcolor{black}{\footnotesize{}95.6} &  & \textcolor{black}{\footnotesize{}-0.01} & \textcolor{black}{\footnotesize{}42} & \textcolor{black}{\footnotesize{}95.2}\tabularnewline
		{\footnotesize{}$\hat{\tau}_{\mathrm{ran}}$} & {\footnotesize{}0.14} & \textcolor{black}{\footnotesize{}26} & \textcolor{black}{\footnotesize{}90.2} &  & \textcolor{black}{\footnotesize{}0.21} & \textcolor{black}{\footnotesize{}14} & \textcolor{black}{\footnotesize{}64.6} &  & \textcolor{black}{\footnotesize{}0.07} & \textcolor{black}{\footnotesize{}37} & {\footnotesize{}94.7}\tabularnewline
		{\footnotesize{}$\hat{\tau}_{\mathrm{cal}}$} & {\footnotesize{}0.01} & {\footnotesize{}26} & {\footnotesize{}94.5} &  & {\footnotesize{}0.02} & {\footnotesize{}11} & {\footnotesize{}95.1} &  & {\footnotesize{}0.00} & {\footnotesize{}33} & {\footnotesize{}95.6}\tabularnewline
		\hline 
		\multicolumn{12}{c}{{\footnotesize{}Scenario 2: Linear outcome \& Probit propensity score}}\tabularnewline
		{\footnotesize{}$\hat{\tau}_{\mathrm{simp}}$} & {\footnotesize{}-0.29} & {\footnotesize{}16} & {\footnotesize{}34.4} &  & {\footnotesize{}-0.08} & {\footnotesize{}9} & \textcolor{black}{\footnotesize{}2.3} &  & {\footnotesize{}-0.22} & {\footnotesize{}30} & {\footnotesize{}65.6}\tabularnewline
		{\footnotesize{}$\hat{\tau}_{\mathrm{fix}}$} & {\footnotesize{}0.08} & {\footnotesize{}35} & {\footnotesize{}90.3} &  & {\footnotesize{}-0.10} & {\footnotesize{}19} & \textcolor{black}{\footnotesize{}4.5} &  & {\footnotesize{}0.12} & {\footnotesize{}69} & {\footnotesize{}90.4}\tabularnewline
		{\footnotesize{}$\hat{\tau}_{\mathrm{ran}}$} & {\footnotesize{}0.24} & {\footnotesize{}28} & {\footnotesize{}73.9} &  & {\footnotesize{}-0.07} & {\footnotesize{}16} & \textcolor{black}{\footnotesize{}29.9} &  & {\footnotesize{}0.21} & {\footnotesize{}60} & {\footnotesize{}85.5}\tabularnewline
		{\footnotesize{}$\hat{\tau}_{\mathrm{cal}}$} & {\footnotesize{}0.01} & {\footnotesize{}22} & {\footnotesize{}94.9} &  & {\footnotesize{}0.01} & {\footnotesize{}11} & \textcolor{black}{\footnotesize{}95.4} &  & {\footnotesize{}0.00} & {\footnotesize{}33} & {\footnotesize{}94.6}\tabularnewline
		\hline 
		\multicolumn{12}{c}{{\footnotesize{}Scenario 3: Linear outcome \& C-loglog propensity
				score}}\tabularnewline
		{\footnotesize{}$\hat{\tau}_{\mathrm{simp}}$} & {\footnotesize{}-0.21} & {\footnotesize{}20} & {\footnotesize{}62.0} &  & {\footnotesize{}-0.21} & {\footnotesize{}10} & {\footnotesize{}41.2} &  & {\footnotesize{}-0.22} & {\footnotesize{}30} & {\footnotesize{}65.6}\tabularnewline
		{\footnotesize{}$\hat{\tau}_{\mathrm{fix}}$} & {\footnotesize{}0.12} & {\footnotesize{}48} & {\footnotesize{}88.8} &  & {\footnotesize{}0.12} & {\footnotesize{}36} & {\footnotesize{}82.7} &  & {\footnotesize{}0.12} & {\footnotesize{}69} & {\footnotesize{}90.4}\tabularnewline
		{\footnotesize{}$\hat{\tau}_{\mathrm{ran}}$} & {\footnotesize{}0.29} & {\footnotesize{}38} & {\footnotesize{}69.1} &  & {\footnotesize{}0.36} & {\footnotesize{}22} & {\footnotesize{}32.5} &  & {\footnotesize{}0.21} & {\footnotesize{}60} & {\footnotesize{}85.5}\tabularnewline
		{\footnotesize{}$\hat{\tau}_{\mathrm{cal}}$} & {\footnotesize{}0.00} & {\footnotesize{}21} & {\footnotesize{}95.3} &  & {\footnotesize{}0.00} & {\footnotesize{}10} & {\footnotesize{}95.1} &  & {\footnotesize{}0.00} & {\footnotesize{}33} & {\footnotesize{}94.6}\tabularnewline
		\hline 
		\multicolumn{12}{c}{{\footnotesize{}Scenario 4: Logistic outcome \& Logistic propensity
				score}}\tabularnewline
		{\footnotesize{}$\hat{\tau}_{\mathrm{simp}}$} & {\footnotesize{}-0.11} & {\footnotesize{}100} & {\footnotesize{}9.1} &  & {\footnotesize{}-0.11} & {\footnotesize{}540} & {\footnotesize{}0.5} &  & {\footnotesize{}-0.11} & {\footnotesize{}160} & {\footnotesize{}20.5 }\tabularnewline
		{\footnotesize{}$\hat{\tau}_{\mathrm{fix}}$} & {\footnotesize{}-0.11} & \textcolor{black}{\footnotesize{}44} & {\footnotesize{}0.3} &  & {\footnotesize{}-0.11} & {\footnotesize{}38} & {\footnotesize{}0.1} &  & {\footnotesize{}-0.11} & {\footnotesize{}39} & {\footnotesize{}0.1}\tabularnewline
		{\footnotesize{}$\hat{\tau}_{\mathrm{ran}}$} & {\footnotesize{}-0.09} & \textcolor{black}{\footnotesize{}33} & {\footnotesize{}1.3 } &  & {\footnotesize{}-0.08} & {\footnotesize{}21} & {\footnotesize{}0.5} &  & {\footnotesize{}-0.10} & {\footnotesize{}34} & {\footnotesize{}0.3}\tabularnewline
		{\footnotesize{}$\hat{\tau}_{\mathrm{cal}}$} & {\footnotesize{}0.01} & {\footnotesize{}74 } & {\footnotesize{}96.3 } &  & {\footnotesize{}0.01} & {\footnotesize{}55} & {\footnotesize{}95.2} &  & {\footnotesize{}0.01} & {\footnotesize{}74} & {\footnotesize{}95.9 }\tabularnewline
		\hline 
		\multicolumn{12}{c}{{\footnotesize{}Scenario 5: Logistic outcome \& Probit propensity
				score}}\tabularnewline
		{\footnotesize{}$\hat{\tau}_{\mathrm{simp}}$} & {\footnotesize{}-0.08} & {\footnotesize{}58} & {\footnotesize{}13.1 } &  & {\footnotesize{}-0.08} & {\footnotesize{}34} & {\footnotesize{}2.3} &  & {\footnotesize{}-0.08} & {\footnotesize{}81} & {\footnotesize{}25.3}\tabularnewline
		{\footnotesize{}$\hat{\tau}_{\mathrm{fix}}$} & {\footnotesize{}-0.10 } & {\footnotesize{}93} & {\footnotesize{}6.9} &  & {\footnotesize{}-0.10 } & {\footnotesize{}85} & {\footnotesize{}4.5} &  & {\footnotesize{}-0.10 } & {\footnotesize{}73} & {\footnotesize{}3.8}\tabularnewline
		{\footnotesize{}$\hat{\tau}_{\mathrm{ran}}$} & {\footnotesize{}-0.08} & {\footnotesize{}67} & {\footnotesize{}23.0 } &  & {\footnotesize{}-0.07} & {\footnotesize{}48} & {\footnotesize{}29.9} &  & {\footnotesize{}-0.09} & {\footnotesize{}61} & {\footnotesize{}8.3}\tabularnewline
		{\footnotesize{}$\hat{\tau}_{\mathrm{cal}}$} & {\footnotesize{}0.01} & {\footnotesize{}89 } & {\footnotesize{}94.7} &  & {\footnotesize{}0.01} & {\footnotesize{}65} & {\footnotesize{}95.4} &  & {\footnotesize{}0.01} & {\footnotesize{}84} & {\footnotesize{}95.0}\tabularnewline
		\hline 
		\multicolumn{12}{c}{{\footnotesize{}Scenario 6: Logistic outcome \& C-loglog propensity
				score}}\tabularnewline
		{\footnotesize{}$\hat{\tau}_{\mathrm{simp}}$} & {\footnotesize{}-0.01} & {\footnotesize{}62} & {\footnotesize{}92.5} &  & {\footnotesize{}-0.01} & {\footnotesize{}34} & {\footnotesize{}92.5} &  & {\footnotesize{}-0.01} & {\footnotesize{}84} & {\footnotesize{}93.0}\tabularnewline
		{\footnotesize{}$\hat{\tau}_{\mathrm{fix}}$} & \textcolor{black}{\footnotesize{}-0.03} & \textcolor{black}{\footnotesize{}53} & \textcolor{black}{\footnotesize{}71.9} &  & {\footnotesize{}-0.03} & {\footnotesize{}50} & {\footnotesize{}69.8} &  & {\footnotesize{}-0.03} & {\footnotesize{}50} & {\footnotesize{}73.7}\tabularnewline
		{\footnotesize{}$\hat{\tau}_{\mathrm{ran}}$} & {\footnotesize{}-0.01} & {\footnotesize{}42} & {\footnotesize{}98.6} &  & {\footnotesize{}0.00} & {\footnotesize{}33} & {\footnotesize{}99.6} &  & {\footnotesize{}-0.02} & {\footnotesize{}44} & {\footnotesize{}92.6}\tabularnewline
		{\footnotesize{}$\hat{\tau}_{\mathrm{cal}}$} & {\footnotesize{}-0.01} & {\footnotesize{}81} & {\footnotesize{}96.0} &  & {\footnotesize{}-0.01} & {\footnotesize{}67} & {\footnotesize{}94.8} &  & {\footnotesize{}-0.01} & {\footnotesize{}82} & {\footnotesize{}94.4}\tabularnewline
		\hline 
	\end{tabular}
\end{table}

\begin{table}
	\caption{\label{tab:Results-3}Balance Check }
	
	\centering{}%
	\begin{tabular}{cccccc}
		\hline 
		&  & simple & fixed  & random & calibration\tabularnewline
		\hline 
		& Cluster 1  & 1.68 & -0.22 & 0.68 & 0.20\tabularnewline
		& Cluster 2 & 1.21 & 0.10 & -0.41 & 0.10\tabularnewline
		$X_{1}$ & Cluster 3  & 1.75 & -0.02 & 0.99 & 0.02\tabularnewline
		& Cluster 4  & 0.86 & -0.04 & -1.05 & 0.02\tabularnewline
		& Cluster 5  & -0.36 & 0.37 & -1.39 & 0.33\tabularnewline
		& Whole Pop  & 1.28 & -0.02 & -0.02 & 0\tabularnewline
		\hline 
		& Cluster 1  & 0.48 & 0.02 & 0.30 & 0.03\tabularnewline
		& Cluster 2 & 0.43 & 0.13 & -0.01 & 0.14\tabularnewline
		$X_{2}$ & Cluster 3  & 0.73 & 0.01 & 0.46 & 0.02\tabularnewline
		& Cluster 4  & 0.18 & -0.08 & -0.34 & -0.07\tabularnewline
		& Cluster 5  & -0.57 & -0.39 & -1.53 & -0.44\tabularnewline
		& Whole Pop  & 0.39 & -0.003 & -0.001 & 0\tabularnewline
		\hline 
	\end{tabular}
\end{table}

\begin{table}
	\caption{\label{tab:Results-2}Results: estimate, variance estimate (ve) based
		on $500$ bootstrap replicates, and $95\%$ confidence interval (c.i.)}
	
	\centering{}%
	\begin{tabular}{cccc}
		\hline 
		& estimate & ve & $95\%$ c.i.\tabularnewline
		\hline 
		simple & 8.78 & 2.11 & (5.94, 11.63)\tabularnewline
		fixed & 0.47 & 0.44 & (-0.83, 1.77)\tabularnewline
		random & 0.52 & 0.44 & (-0.77, 1.82)\tabularnewline
		calibration & 0.53 & 0.39 & (-0.71, 1.76)\tabularnewline
		\hline 
	\end{tabular}
\end{table}

\end{document}